\documentclass[twocolumn,showpacs,preprintnumbers,amsmath,amssymb]{revtex4}

\usepackage{graphicx}
\usepackage{dcolumn}
\usepackage{bm}

\begin{document}


\title{Vortex dynamics in two-dimensional Josephson junction arrays}

\author{\large{Md. Ashrafuzzaman, Massimiliano Capezzali$^{*}$ and Hans Beck }}
\affiliation{%
Institut de Physique, Universit\'e de Neuch\^atel, Rue A.L.Breguet, 2000 Neuch\^atel, Switzerland\\
*Cellular Biophysics and Biomechanics Laboratory, Swiss Federal Institute of Technology, 1015 Ecublens, Switzerland\\
Contact: md.ashrafuzzaman@unine.ch, hans.beck@unine.ch
}%


\begin{abstract}
The dynamic response of unfrustrated two-dimensional Josephson junction arrays close to, but above the Kosterlitz-Thouless($KT$)
       transition temperature is described in terms of the vortex dielectric function $\epsilon(\omega)$. The latter is calculated by
       considering separately the contribution of $\ll$free$\gg$ vortices interacting by a screened Coulomb potential, and the $\ll$pair
       motion$\gg$ of vortices that are closer to each other than the $KT$ correlation length. This procedure allows to understand
       various anomalous features in $\epsilon(\omega)$ and in the flux noise spectra that have been observed experimentally and in dynamic
       simulations.  
\end{abstract}

\maketitle
Two-dimensional ($2d$) Josephson junction arrays ($JJA$) are described by the classical $XY$-model in which the superconducting phases $\theta_l$ of neighboring sites $l$ are coupled by the Josephson interaction. The relevant collective phase excitations are vortices ($V$) and antivortices ($A$). They behave as a $2d$ neutral Coulomb gas ($CG$). Their charge is related to the Josephson coupling $J$ by $q_o^2 = 2\pi J$. At the Berezinsky-Kosterlitz-Thouless ($BKT$) transition the $CG$ goes over from a $\ll$dielectric$\gg$ phase consisting of $VA$-pairs to a $\ll$metallic$\gg$ phase containing free $V$ and $A$.  One of its main signatures is the universal jump of the helicity modulus at $T_{BKT}$. It has been observed experimentally, both in the current-voltage characteristics of the array and in its response to an electromagnetic field of very low frequency [1,2].

The dynamics of $JJAs$ is less well understood. For the equations of motion for the superconducting phases the $\ll$resistively shunted junction model$\gg$ ($RSJ$) can be used for arrays in which electrostatic charging effects may be neglected. Based on this model an equation of motion for the vortex excitations can be derived [3],
describing them as massless point particles, subject to a friction force coming from normal current losses and interacting via the $2d$
Coulomb interaction, varying essentially with the logarithm of their distance. The dynamics of such a system is usually treated by
considering separately the motion of bound pairs and of free particles. The simplest picture for free motion is given by the Drude form of the dynamic dielectric function $\epsilon(\omega)$, expressed in terms of the friction constant. Minnhagen [4]
has developed a more sophisticated expression for $\epsilon(\omega)$ by assuming that this quantity can be derived from the static wave number dependent
dielectric function, taken to have a Debye screening form, by replacing the wave number by frequency. The result - usually referred to as $\ll$Minnhagen
phenomenology$\gg$ ($MP$) - differs in an essential way from Drude's ($D$) behaviour. In particular $Re(1/\epsilon_D(\omega))\propto \omega^2$,
whereas $Re(1/\epsilon_{MP}(\omega))\propto \arrowvert\omega\arrowvert$, and the so-called $\ll$peak-ratio$\gg$ r, given by
\begin{equation}
r=\frac{Im(\frac{1}{\epsilon(\omega_{max})})}{Re(\frac{1}{\epsilon(\omega_{max})})}
\end{equation}
where $\omega_{max}$ is the frequency at which $Im(1/\epsilon)$ has its maximum, is $r_D = 1$, whereas $r_{MP} = 2/\pi$ is smaller.

The dynamics below $T_{BKT}$ is usually treated by averaging the dynamic response of a pair of separation $d$ over a probability distribution
for $d$ [5,6]. 

Three main types of experiments aim at elucidating dynamical properties of $JJAs$. The exponent of the non-linear current-voltage
characteristics is related to the dynamic critical exponent describing the critical slowing down of the $2d$ $CG$ near the $BKT$ transition
(see [1]). The dynamic conductance $G(\omega)$ of the array can be inferred from measuring the dynamic response of the array to a time dependent
current in a two-coil experiment [1,2,7]. Measurements on frustrated arrays [8] have produced results that are closer to the $MP$ prediction than to the simple Drude form,
in particular as far as the above-mentioned peak-ratio and the frequency dependence of 1/$\epsilon(\omega)$ are concerned:
$Re(1/\epsilon(\omega))\propto\arrowvert\omega\arrowvert$  over a
sizable range of frequencies, which is a signature of anomalous $MP$ dynamics. Similar results have been obtained by analytical
calculations [6,9,10].

Measuring the temporal fluctuations of the magnetic flux through a given area of a $JJA$ [11,12] gives another insight into the dynamics
of the currents in the array, and thus of the vortex system. At sufficiently low frequencies, above $T_{BKT}$, the time Fourier transform
$S_{\phi}(\omega)$of the noise is white (frequency independent). For larger $\omega$ these experiments show a rather extended region where
$S_{\phi}(\omega)\propto 1/\omega$,
which is unexpected and needs a deeper explanation.

Various analytical and numerical methods have been used aiming at explaining the somewhat unexpected $\ll$anomalous$\gg$ dynamics of $JJAs$
revealed by the above mentioned experiments. Dynamic simulations for $JJAs$ [13-16] have been performed based on the equations of motion
for the phases of the array or for the $CG$ [17]. The extensive work of Minnhagen and collaborators has confirmed the existence of a frequency interval in which the anomalous $MP$ dynamics
can be seen, both for zero field and for frustrated arrays. On the other hand the flux noise spectrum obtained in these and other
calculations usually only show a common $1/\omega$ tangent to the curves for different temperatures, but no extended region with
$1/\omega$-noise. For higher frequencies $S_{\phi}(\omega)\propto 1/\omega^{3/2}$ [18-20], which is characteristic of vortex diffusion or
$S_{\phi}(\omega)\propto 1/\omega^{2}$, for
even higher frequencies [20]. Other calculations do yield $1/\omega$-noise [21] or $S_{\phi}(\omega)\propto 1/\omega^{a(T)}$ with an exponent $a(T)$ close
 to $1$, but slowly varying with temperature [22].

Our theoretical approach to $JJA$ dynamics is based on the equations of motion for the $2d$ neutral $CG$ containing a friction term and the
effect of the Coulomb interaction. The key quantity for calculating all observables of interest is the dynamic charge correlator:
\begin{equation}
\phi_{\rho\rho}({\bf{k}},z)=\int_0^{\infty}dte^{-zt}\langle\rho({\bf{k}},t)\rho^*({\bf{k}})\rangle
\end{equation}
$\rho({\bf{k}})$ being the Fourier transform of the charge density. The equal time correlation function $S(k)=\langle\arrowvert\rho({\bf{k}})\arrowvert^2\rangle$ is the charge structure factor of the $CG$. From $\phi_{\rho\rho}$ one can obtain the dynamic charge susceptibility and the dielectric function of the $CG$ [25]. The later can be expressed in terms of a dynamic vortex mobility $\mu$
\begin{equation}
\epsilon(\omega)=1-\frac{2\pi q_o^2\mu(\omega)}{i\omega}
\end{equation} 
and it yields the superconducting part of the dynamic conductance of the $JJA$:
\begin{equation}
G_s(\omega)=\frac{J}{i\omega\epsilon(\omega)}
\end{equation}


\begin{figure} [h]
\let\picnaturalsize=N
\def\picsize{7 cm}
\def\picfilename{DFLinLinReImMori.eps}
\ifx\nopictures Y\else{\ifx\epsfloaded Y\else\input epsf \fi
\let\epsfloaded=Y
\centerline{\ifx\picnaturalsize N\epsfxsize \picsize\fi \epsfbox{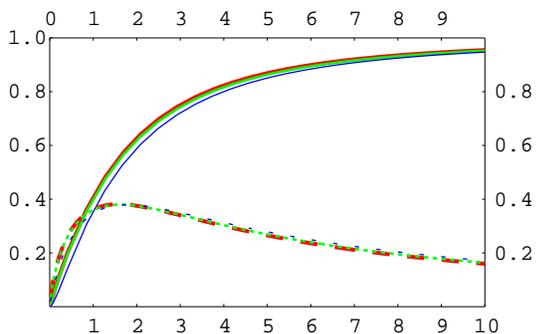}}}\fi
\caption{{\bf $Re[1/\epsilon(\omega)]$  and $-Im[1/\epsilon(\omega)]$ vs
$\omega/\omega_a$ plot (dotted curves for $-Im$) for free $V$ and $A$. From more thickness to less for $T/T_{BKT}$= 1.01, 1.05, 1.10.}}
\label{AmplitudePhi4}
\end{figure}


Assuming that supercurrent is the main source of the flux threading through a coil of radius $R$ at a distance $d$ from the array plane, $S_{\phi}(\omega)$ is related to $\phi_{\rho\rho}$:
\begin{eqnarray}
S_{\phi}(\omega)=&&2(2\pi^2R\mu_o)^2(\frac{2e}{\hbar})^2(J\lambda)^2\times{}
\nonumber\\
&&{} \int_0^{\infty}dk\frac{J_1(kR)^2e^{-2kd}}{k(1+\lambda k)^2}Re\phi_{\rho\rho}(k,-i\omega)
\end{eqnarray}
$J_1$ being the first order Bessel function and $\lambda$ the magnetic penetration depth of the $JJA$.

Our goal consists in elucidating - by analytical means - the critical dynamics of the $VA$-system when $T_{BKT}$ is approached from above.
In this regime two types of motion, occurring at different length scales, have to be considered simultaneously. Particles closer together
than a typical $T$-dependent critical distance $d_c$ are supposed to move as a bound pair, at least for times smaller than some typical life
time $\tau$ (depending also on $T$). They interact through the Coulomb potential, screened as in a dielectric. On the other hand,
particles further apart from each other than $d_c$ will be supposed to move like $\ll$free$\gg$ - i.e. $\ll$unbound$\gg$ - particles, subject to a
metallically screened potential. A complete theoretical approach should, of course, treat the two types of time evolution with the
same mathematical tools. Here we have chosen to evaluate separately the mobilities $\mu_b$ for $\ll$bound$\gg$ and $\mu_f$ for $\ll$free$\gg$ motion. The two
results are then combined by calculating the total mobility as a weighted sum of $\mu_b$ and $\mu_f$ , the relative weights being given by the
density of paired and free excitations.

For the free motion $\mu_f(\omega)$ is found by applying Mori's technique of calculating dynamic correlation functions to $\phi_{\rho\rho}$ [6,10]. As usual higher order correlators, showing up when the equations of motion
for the individual particles are used, are factorized. In order that this factorization be adequate for a system with long range
interaction, we replace the Coulomb potential by its screened version. Its Fourier transform has the form
\begin{equation}
V_{sc}({\bf{p}})=\frac{q_o^2}{p^2+\xi_{sc}^{-2}}
\end{equation}
with $\xi_{sc}$ being the (metallic) screening length which we identify with $KT$ correlation length. This replacement is standard in the
treatment of quantum Coulomb systems [23]. The resulting density correlator (2) has the usual form
\begin{equation}
\phi_{\rho\rho}({\bf{k}},z)=\frac{S({\bf{k}})}{z+\frac{k_BTk^2}{S({\bf{k}})\gamma(z)}}
\end{equation}
The generalized friction function $\gamma(z)=\mu^{-1}(z)$ is the memory kernel of the charge current density. $\gamma_f(z)$ for free motion shows $MP$ behavior over a frequency range which increases when $T_{BKT}$ is approached. There are three
frequency regimes, separated by the scale frequencies $\omega_a=\frac{k_BT}{\Gamma a^2}$ with $a$ being the lattice constant and $\omega_{\xi}=\frac{k_BT}{\Gamma\xi^2}$. $Re[\gamma_f(-i\omega)]$ is flat for
$\omega<\omega_{\xi}$, it coincides with the bare friction parameter $\Gamma$ for $\omega>\omega_a$, whereas in between  $Re(\gamma_f)\propto\ln\arrowvert \omega \arrowvert$. This is precisely what $MP$ predicts, and the corresponding inverse dielectric function shown
in figure 1 indeed follows the $MP$ prediction for a frequency range that increases when $T_{BKT}$ is approached. The values of the peak ratio
(1), varying between 0.67 and 0.73 for the three temperatures shown in figure 1, are also much closer to $MP$ than to Drude. Thus, in this framework the
anomalous $MP$ dynamics is explained by the increasing influence of the long range Coulomb force,
which makes the motion more and more $\ll$sluggish$\gg$. Whereas at high temperatures the potential is (metallically) screened for all relevant length scales, this screening becomes less and
less efficient when $T_{BKT}$ is approached where the screening length $\xi_{sc}$ diverges. However, contrary to $MP$ the response always crosses over
to Drude-like when $\omega<\omega_{\xi}$. If this were not the case the arrays would still be superconducting above $T_{BKT}$ [10], whereas in reality a
finite flux-flow resistance yields a finite conductance $G(\omega=0)$. The corresponding flux noise spectrum, however, does not show any
extended $1/\omega$ region. Both results, 1/$\epsilon(\omega)$ and $S_{\phi}(\omega)$, are very similar to Minnhagen's simulations. One may thus draw the
conclusion that anomalous $MP$ behaviour for the dielectric function does not yield the - equally $\ll$anomalous$\gg$ - $1/\omega$ noise, which should thus have a different origin.
\begin{figure} [h]
\let\picnaturalsize=N
\def\picsize{7 cm}
\def\picfilename{fluxnoisecdipoleAHNSnewo.eps}
\ifx\nopictures Y\else{\ifx\epsfloaded Y\else\input epsf \fi
\let\epsfloaded=Y
\centerline{\ifx\picnaturalsize N\epsfxsize \picsize\fi \epsfbox{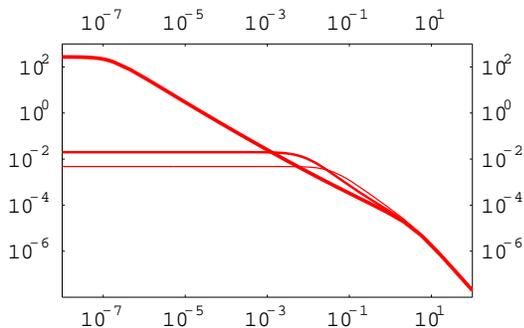}}}\fi
\caption{{\bf $S_{\phi}(\omega)$ vs $\omega/\omega_a$ plot for $VA$ pair dynamics in method (b) with fixed $k_{o1}^2=1.0$. From more thickness
to less for $T/T_{BKT}$= 1.01,1.05,1.10.}}
\label{AmplitudePhi4}
\end{figure}
\begin{figure} [h]
\let\picnaturalsize=N
\def\picsize{7 cm}
\def\picfilename{fluxnoiseboundAHNSold.eps}
\ifx\nopictures Y\else{\ifx\epsfloaded Y\else\input epsf \fi
\let\epsfloaded=Y
\centerline{\ifx\picnaturalsize N\epsfxsize \picsize\fi \epsfbox{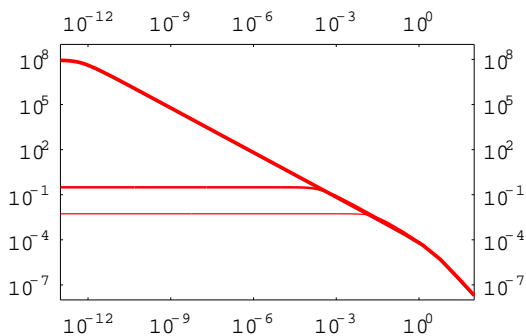}}}\fi
\caption{{\bf $S_{\phi}(\omega)$ vs $\omega/\omega_a$ plot  with $T$ dependent $k_{o2}^2$ for $VA$ pair dynamics in method (a). From more thickness
to less for $T/T_{BKT}$= 1.01,1.05,1.10.}}
\label{AmplitudePhi4}
\end{figure}

Turning now to the $\ll$paired$\gg$ motion we average the dynamic polarizability of a single pair over a suitable
probability distribution function $\phi(r)$ for the pair size $r$ [5,6]:
\begin{equation}
\chi(z)=\int_a^{\infty}rdr\frac{\phi(r)}{z\Gamma+f(r)+\frac{1}{\tau(r)}}
\end{equation}
We have chosen a normalized Boltzmann factor for $\phi(r)$
\begin{equation}
\phi(r)\propto e^{-\beta\frac{q_o^2\ln(r/a)}{\epsilon(r)}}\theta(d_c-r)
\end{equation}
using two different forms for the static dielectric function screening the bare Coulomb potential ; (a) $\epsilon(r)$ fixed at $\epsilon(\infty,T_{BKT})=\pi J/(2k_BT_{BKT})$ and (b) $\epsilon(r)$ length scale dependent, as determined in the framedwork of the $KT$ scaling equations [4]. The distribution is cut off at the
length scale $d_c$. For (a) this is the $BKT$ correlation length $\xi_{BKT}$, and for (b) it is the distance where the screened potential turns over
from attractive to repulsive. Both have the characteristic $BKT$ temperature dependence with simply somewhat different parameters. The two procedures (a) and (b) yield similar
results. In the same spirit the force constant in the denominator of (8) is given by
\begin{equation}
f(r)=\frac{q_o^2}{r^2\epsilon(r)}
\end{equation}
We have also introduced a finite pair life time [6] given by
\begin{equation}
\frac{1}{\tau(r)}=\omega_o \exp\{\frac{q_o^2\ln(d_c/a)}{\epsilon(d_c)}-\frac{q_o^2\ln(r/a)}{\epsilon(r)}\}
\end{equation}
where $\omega_o$, a free parameter, is an $\ll$attempt frequency$\gg$ for pairs trying to escape the barrier up to the maximum of the potential at the
distance $d_c$ by thermal excitation. The inverse dielectric function resulting from (8) shows again  marked deviations from Drude
behaviour, in particular for $Re(1/\epsilon(\omega))$ varying like $\omega^{s(T)}$ for intermediate $\omega$. The $T$-dependent exponent $s(T)\approx$ 1/6 to 1/3 is due to the $r$-integral (8) where the dependence of $\phi$, $f$ and $1/\tau$ on $r$ imposes a particular exponent to the frequency dependence of the susceptibility.
This even $\ll$more anomalous$\gg$ frequency dependence is reflected in the peak ratio $r$ (equation (1)) which is now even smaller than the $MP$
value of $2/\pi$. The low $\omega$ cross-over to Drude behaviour occurs at a critical frequency $\omega_c$, which is found to, vary as $d_c^{-2}$, whereas for $\omega>\omega_a$
it crosses over to the high frequency Drude form. In the window $\omega_c<\omega<\omega_a$ the behaviour is anomalous. This has very
interesting consequences for the flux noise spectrum presented in figures 2 and 3. The values used for the parameters showing up in the integral (8)
are given in the figure caption. The structure factor in the density correlator (7) has been given the following form
\begin{equation}
  S(k)=\frac{k^2}{k^2+k_{o}^2}
\end{equation}
which respects the charge neutrality of the $V-A$ system and the correct limit for large $k$. Two different ideas have been used for the inverse characteristic length $k_o$ in $S(k)$. For $2d$ neutral Coulomb gas it is simply related to the charge $q_o$ and the density $n$ of the particles: $k_{o1}^2=\frac{k_BT}{2\pi q_o^2n}$. Alternatively, for evaluating the contribution to flux noise of pairs of particles subject to bound motion one may relate $k_o$ to the
mean distance of such pairs, resulting from the distribution (9): $k_{o2}^2=\frac{\pi^2}{\langle r^2\rangle}$. This corresponds to the small-$k$ form of $S(k)$ for independent pairs. Choosing $k_{o1}$ yields a flux noise spectrum (figure 2) which behaves as $1/\omega^{a(T)}$, where $a(T)=1$ at  $T_{BKT}$ and increases above. The curves for different temperatures cross each other (there is a hint that this happens in the data of reference [12]). On the other hand (figure 3) this temperature dependence of the exponent (coming from the $T$-dependence of the integrand in (8)) is compensated by using the $T$-dependent values of $k_{o2}$ in (5), see figure 3. This approach emphasizes the fact that large pairs, for which $f(r)$ and $1/\tau(r)$ are small, give the main contribution to (8) and it is the combined
effect of the exponent $a(T)$ and the structure factor that leads to a $1/\omega$ flux noise, and all the curves for different temperatures
fall on top of each other for $\omega>\omega_c$, as it is seen in the experiments [11,12]. The white noise level showing up for $\omega<\omega_c$ is
also strongly $T$-dependent. According to (5) and (7) it is given by
\begin{equation}
S_{\phi}(0)=\frac{1}{(2\pi)^4}\frac{\gamma(\omega=0)}{k_BTn}(\frac{k_BT}{J})^2
\end{equation}

\begin{figure} [h]
\let\picnaturalsize=N
\def\picsize{7 cm}
\def\picfilename{ReImliliDFnAHNSnewbfPLT.eps}
\ifx\nopictures Y\else{\ifx\epsfloaded Y\else\input epsf \fi
\let\epsfloaded=Y
\centerline{\ifx\picnaturalsize N\epsfxsize \picsize\fi \epsfbox{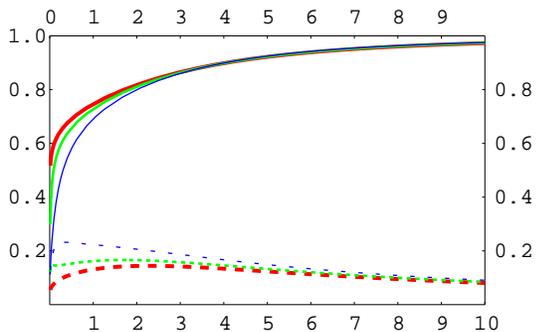}}}\fi
\caption{{\bf $Re[1/\epsilon(\omega)]$ and $-Im[1/\epsilon(\omega)]$ vs $\omega/\omega_a$ plot (dotted curves for $-Im$) using the combined mobility for pair and free motion. From more thickness to less for $T/T_{BKT}$= 1.01,1.05,1.10.}}
\label{AmplitudePhi4}
\end{figure}

The dominant contribution comes from $\gamma(0)$ and thus $S_{\phi}(\omega)\sim d_c^2$. However, when $d_c$
reaches the sample size $L$, $\gamma(0)$ saturates and the $T$-dependence of $S_{\phi}(0)$ comes from the remaining factors in (13), in particular the
density n of $VA$-excitations. This $\ll$masking$\gg$ of the true critical slowing down has been observed in Ref. [12].

As a final step we combine the two contributions to obtain the total mobility $\mu(\omega)$: $\mu(\omega)=(1-\nu_f)\mu_b(\omega)+\nu_f\mu_f(\omega)$, with $\nu_f=\frac{1}{(na^2)(d_{c}/a)^2}$. Bound motion is still dominant up to $T/T_{BKT}$ = 1.1, since its weight is larger, although the free mobility
itself exceeds the one of bound pairs. Combining the two contributions (figure 4) yields a rather extended
flat region of Im(1/$\epsilon(\omega)$) for intermediate frequencies, which is another signature of $\ll$anomalous$\gg$ vortex dynamics found also in
numerical simulations[24].

In conclusion our approximate analytical calculations, separating $\ll$free$\gg$ and $\ll$paired$\gg$ motion, give the following insight into vortex
dynamics in $JJAs$ above the Berezinski-Kosterlitz-Thouless temperature :

      $\bullet$ The anomalous Minnhagen phenomenology ($MP$) is a consquence of the motion of (unbound) vortices in a Coulomb potential which is
       screened by other $\ll$free$\gg$ particles. The more this screening is reduced when approaching the transition temperature, the more $MP$
       is pronounced. This regime does not lead to $1/\omega$ flux noise.

       $\bullet$ $V$ and $A$ moving - at short enough distances and up to some finite life time - as a pair yield an even more anomalous vortex
       dielectric constant with temperature dependent frequency exponents. This effect, combined with a temperature dependent pair
       structure factor produces $1/\omega$ flux noise in an intermediate frequency range.\\

We thank P. Martinoli, D. Bormann and S. Korshunov for interesting discussions. This work was supported by the Swiss National Science
Foundation. Md.A. thanks the Swiss Confederation for his scholarship.

\bf{References}\\
\begin{scriptsize}
[1] for a summary of experimental work see :R.S.Newrock, C.J.Lobb,U.Geigenmüller, M.Octavio ; Solid State Physics 54, 263 (2000)

[2] P.Martinoli. C.Leemann ; Journal of Low Temperature Physics 118, 699 (2000)

[3] H.Beck, D.Ariosa ; Solid State Communications 80, 657 (1991)

[4] P.Minnhagen ; Rev Mod Phys 59, 1001 (1987)

[5] V.Ambegaokar, B.I.Halperin, D.R.Nelson, E.D.Siggia; Phys Rev Lett 40, 783 (1978) and Phys Rev B21, 1806 (1980)

[6] D.Bormann, H.Beck, O.Gallus, M.Capezzali ; J Phys IV France 10, Pr5-447 (2000)

[7] P.Martinoli, Ph.Lerch, Ch.Leemann, H.Beck ; Japanese Journal of Applied Physics 26-3, 1999 (1987)

[8] R.Th$\acute{e}$ron, J.B.Simond, C.Leemann, H.Beck, P.Martinoli, P.Minnhagen ; Phys Rev Lett 71, 1246 (1993)

[9]  S.E.Korshunov;Phys Rev B 50, 13616(1994); H.Beck ; Phys Rev B 49, 6153 (1994)

[10]  M. Capezzali, H. Beck, S.R.Shenoy ; Phys Rev Lett 78, 523 (1997)

[11] T.J.Shaw, M.J.Ferrari,L.L.Sohn, D.H.Lee, M.Tinkham, J.Clarke ; Phs Rev Lett 76, 2551 (1996)

[12] S.Candia, Ch.Leemann, S.Mouaziz, P.Martinoli ; cond-mat/0111212, to appear in Physica C

[13] P.H.E.Tiesinga, T.J.Hagenaars, J.E.van Himbergen, J.V.Jose ; Phys Rev Lett 78, 519 (1997)

[14] I-J Hwang, D.Stroud ; Phys Rev B 57, 6036 (1998)

[15] L.M.Jensen, B.J.Kim, P. Minnhagen ; cond-mat/0003447 ; Phys Rev B 61,15412 (2000)

(this article cites numerous other references to work done by the same group)

[16] Md. Ashrafuzzaman, H. Beck ; $\ll$Studies of High Temperature Superconductors$\gg$, Nova Science Publishers Inc., New York, Vol 43 (2002), Ch.5; cond-mat/0207572

[17] B.J.Kim, P.Minnhagen and references cited there in ; cond-mat/0011297 (Nov 2000)

[18]  Carsten Timm ; Phys Rev B 55, 3241 (1997)

[19]  Ing-Jye Hwang, D.Stroud ; Phys Rev B 57, 6036 (1998)

[20]  Beom Jun Kim, P. Minnhagen ; Phys Rev B 60, 6834(1999)

[21]  K-H Wagenblast, R. Fazio ; cond-mat/9611177

[22] P.H.E. Tiesinga, T.J.Hagenaars, J.E. van Himbergen, J.V.Jose ; Phys Rev Lett 78, 519 (1997)

[23] H.E.DeWitt, M.Schlanges, A.Y.Sakakura, W.D.Kraeft ; Phys Lett A197, 326 (1995)

[24] A. Jonsson, P. Minnhagen ; Phys Rev B 55, 9035 (1997)

[25] N.H. March, M.P. Tosi; $\ll$Atomic Dynamics in Liquids$\gg$, John Wiley and Sons, New York, 1976; ch.9
\end{scriptsize}

\end{document}